\documentclass[bibyear]{aa} % if the references are not structured 
\usepackage{graphicx}
\usepackage{amsmath}
\usepackage{float} % para usar [H]
\usepackage{caption}
\usepackage{xcolor}
\usepackage{soul}

\usepackage{txfonts}
\begin{document}

  \title{Doppler beaming factors for white dwarfs, main sequence stars, and giant stars}
 { }
   \subtitle{Limb-darkening coefficients for 3D (DA and DB) white dwarf models}

\author{A. Claret \inst{1, 2} \and E. Cukanovaite \inst{3}, K. Burdge \inst{4}, 
        P.-E. Tremblay\inst{3}, S. Parsons \inst{5} \and T. R. Marsh\inst{3}}
   \offprints{A. Claret, e-mail:claret@iaa.es. Tables 1-34 are   available 
in electronic form at the CDS via anonymous ftp or directly from the authors. }
\institute{Instituto de Astrof\'{\i}sica de Andaluc\'{\i}a, CSIC, Apartado 3004,
            18080 Granada, Spain
            \and
            Dept. F\'{\i}sica Te\'{o}rica y del Cosmos, Universidad de Granada, 
            Campus de Fuentenueva s/n,  10871, Granada, Spain\
            \and 
             Department of Physics, University of Warwick, Coventry CV4 7AL, UK 
             \and
            Division  of Physics, Mathematics and Astronomy, California Institute 
            of Technology, Pasadena, CA 91125, USA
             \and
            Department of Physics and Astronomy, University of Sheffield, Sheffield S3 7RH, UK}
            \date{Received; accepted; }

\abstract
{Systematic theoretical calculations of Doppler beaming  factors are scarce in the literature, 
particularly in the case of white dwarfs. Additionally, there are no specific calculations 
for the limb-darkening coefficients of 3D white dwarf models.}
{The objective of this research is to provide the astronomical community with Doppler 
beaming calculations for a wide range of effective temperatures, local gravities, and hydrogen/metal 
content for white dwarfs as well as stars on both the main sequence and the giant branch.
In addition, we present the theoretical calculations of the limb-darkening 
coefficients for 3D white dwarf models for the first time.}
{We computed Doppler beaming  factors for DA, DB, and DBA white dwarf models, as well as for main sequence and 
giant stars covering  the transmission curves of the Sloan, UBVRI, HiPERCAM, Kepler, TESS, and 
Gaia photometric systems. The calculations of the limb-darkening coefficients for 3D models 
were carried out  using the least-squares method  for these photometric systems. }
{The input physics of the white dwarf models for which we have computed the Doppler 
beaming  factors are:  chemical compositions $\log$ [H/He] = $-$10.0 (DB), $-$2.0 (DBA), 
and He/H = 0 (DA), with $\log g$ varying between 5.0 and 9.5 and effective temperatures 
in the range 3750-100\,000~K. The beaming  factors were also calculated assuming non-local 
thermodynamic equilibrium (NLTE) for the case of DA white dwarfs with T$_{\rm eff}$ > 40\,000~K. 
For the mixing-length parameters, 
we adopted      ML2/$\alpha$ = 0.8  (DA case) and 1.25 (DB and DBA). 
The Doppler beaming  factors for 
main sequence and giant stars were computed using the ATLAS9 version, characterized   
by  metallicities ranging from [-2.5, 0.2] solar abundances, with $\log g$ varying between 0 
and 5.0 and effective temperatures between 3500 and 50\,000 K. 
The adopted  microturbulent velocity for these models was  2.0 km s$^{-1}$. The limb-darkening 
coefficients were computed for three-dimensional DA and DB white dwarf models calculated with the CO$^5$BOLD 
radiation-hydrodynamics code. The parameter range covered by the three-dimensional DA models spans $\log{g}$ values 
between 7.0 and 9.0 and $T_{\rm{eff}}$  between 6000 and 15000 K, while He/H = 0.  The three-dimensional DB models cover 
a similar parameter range of $\log{g}$ between 7.5 and 9.0 and $T_{\rm{eff}}$ between 12\,000 
and 34\,000 K, while $\log{\mathrm{H/He}} = -10.0$.
We adopted six laws for the computation of the limb-darkening coefficients: linear, 
quadratic, square root, logarithmic, power-2, and a general one with four 
coefficients. 
}
{The beaming  factor calculations, which use realistic models of stellar atmospheres, show
that the black body approximation is not accurate, particularly for the filters $u$, $u'$, $U$, $g$, $g'$, and $B$. 
The black body approach is only valid for high effective temperatures and/or long  effective wavelengths. 
Therefore, for more accurate analyses of light curves, we recommend the use of the beaming  factors presented in this paper. 
Concerning limb-darkening, the distribution of specific intensities for 3D models indicates that, in general,  these 
models are less bright toward the limb than their 1D counterparts, which implies steeper 
profiles. To describe these intensities better, we recommend  
the use of the  four-term law (also for 1D models) given the level of precision that is being achieved 
with Earth-based instruments and space missions such as Kepler and TESS (and PLATO in the future). 
}

   \titlerunning {Doppler beaming factor for white dwarfs}
   \maketitle
%
%________________________________________________________________

\section{Introduction}

Some time ago, Loeb \& Gaudi (2003) indicated the possibility of detecting exoplanets by 
using the Doppler beaming technique, which causes an asymmetry in the ellipsoidal modulation in 
the light curves of binary systems. This technique has an additional 
advantage because, unlike the transit method, it can be applied 
to practically any orbital inclination of the system.

The contribution of beaming is essential to synthesizing the light curves of a given  
system under study. However, such calculations are sparse in the literature. 
Bloemen (2015) has made an important effort in this regard, but his calculations are 
limited to a few systems. One of the main objectives of this short paper is to provide users with  the calculations of the Doppler beaming  factors for white dwarfs, stars on the main 
sequence, and stars on the giant branch. Such calculations cover a wide range 
of effective temperatures, $\log g$, and hydrogen/metal content, and they are available for 
some of  the most commonly used photometric systems: Sloan, UBVRI, Kepler, TESS, Gaia, and HiPERCAM.

On the other hand, some studies have shown that 3D atmosphere models compare better with observations 
of the  center to limb variation for the Sun 
(see Pereira et al.~2013). Pereira et al.~(2013) find a very good agreement between the limb-darkening 
from 3D models and observations at ultraviolet and visible wavelengths. The  observations used in 
their paper were those of Pierce \& Slaughter (1977) and Neckel \& Labs (1994). However, 
Pereira et al. (2013)  also find a poorer agreement for longer wavelengths (between 12\,000 and 
24\,000\,$\textup{\r{A}}$). In order to test the validity of stellar atmosphere models in other kind of stars, 
we present a series of computations of limb-darkening coefficients (LDCs) for white dwarfs of spectral 
types DA and DB, based on 3D model atmospheres and covering the same range of photometric systems as mentioned above. 

We organize the paper as follows:  Sect. 2 is dedicated to the computation of the Doppler beaming  factors 
while Sect. 3 is devoted to the LDCs of DA-3D and DB-3D  white dwarf models. 
Finally, in Appendix A we give brief explanations of Tables~A1 and~A2 and the 
data they contain. Additional material or calculations for specific photometric systems can be carried
 out upon request.

\section{Doppler beaming factors for DA, DB, and DBA white dwarfs,  main sequence stars, and giant stars}

Doppler beaming is caused by the radial velocity of the stars in a double system, which displace the 
spectrum and modulate the photon emission rate toward the observer. The first theoretical investigation into this effect 
was carried out by Hills \& Dale (1974) in the case of rotating white dwarfs and 
by Shakura \& Postnov (1987) for the orbital movement of double systems. 

 From the observational point of view, there have also been important
        advances related to beaming effects. We can mention, for example:
        Mazeh \& Faigler (2010), who detected an ellipsoidal modulation
        and relativistic beaming effect in the exo-planetary system CoRot-3; and
        Ehrenreich et al. (2011), who, using spectroscopic techniques, confirmed
        the relativistic beaming effect in the KOI-74 system consisting of a
        hot compact object around an early-type star. More recently, Wong et al.
        (2020) performed a consistent analysis of the KOI-964 system, which consists of a hot white dwarf and an A-type host star. With respect to the Doppler
        beaming factor, Wong et al. (2020) find a good agreement between theory
        and observational data.

In a previous paper on gravity and LDCs (Claret et al.~2020), we computed the beaming factors for four white dwarf  models - DA, DB, DBA, and  
non-local thermodynamic equilibrium (NLTE) DA - adopting the black body approach. However, such a simple  method neglects 
the effects of the Balmer jump, of the absorption lines, and even of the shape of 
the spectra, which is different from black body radiation. Furthermore, such an approach does not predict 
any dependence on $\log g$ or atmospheric composition. 
As we will see below, the local 
gravity and the hydrogen content (or metallicity in the case of stars) may influence the 
calculation of the beaming factors.
 
As mentioned in the Introduction, there are no systematic calculations for beaming 
factors that consider realistic atmosphere models of white dwarfs or stars on 
the main sequence or the giant branch. In the present paper, we adopt white dwarf   
atmosphere models of the types DA, DB, DBA, and DA-NLTE as well as the ATLAS9 
models (main sequence, subgiants, and giants) for the calculation of the beaming factors  
for several passbands. The description of the adopted white dwarf models 
(DA, DB, DBA, and DA-NLTE) can be found in Claret et al.~(2020, Sect. 2).
The models for stars on the main sequence, subgiants, and giants used here were 
calculated with the ATLAS9 version.  The computation of beaming factors are presented for  metallicities 
ranging from [-2.5, 0.2] solar abundances, with $\log g$ varying between 0 and 5.0 
and effective temperatures between 3500~K and 50\,000~K. The adopted  microturbulent 
velocity is  2.0 km s$^{-1}$. For more detailed information on ATLAS9 features, see, for example, Castelli et al.~(1997). 
 
If we assume that the monochromatic flux is proportional to a frequency power, that is to say
F($\nu) \propto \nu^{\alpha}$ (Shakura \& Postnov 1987),  for the flux variation due 
to the Doppler shift (Loeb \& Gaudi 2003)  we have

\begin{eqnarray}
 {F(\nu)_{\rm obs}} = {F(\nu)_0} \left(1+(3-\alpha){v_{\rm r}\over{c}}\right),
\end{eqnarray}

\noindent
where $v_{\rm r}$ is the radial velocity (nonrelativistic approach), $c$ is the velocity of light in vacuum, 
$F(\nu)_{\rm obs}$ is the observed flux, and $F(\nu)_0$ is the monochromatic 
flux  in the absence of the source motion.    Equation 1  can also be written as 

\begin{eqnarray}
    F(\lambda)_{\rm obs} = F(\lambda)_0    \left( 1 - B(\lambda) \frac{v_r}{c}\right), 
\end{eqnarray} 

\noindent
where  $\lambda$ is the wavelength and  B($\lambda$) is the monochromatic beaming factor,  

\begin{eqnarray}
B(\lambda)= 5 + {{{\rm d}\ln F(\lambda)}\over{{\rm d}\ln \lambda}}.  
\end{eqnarray}

To compute the  photon weighted bandpass-integrated beaming factors, we used the following equation:  

\begin{eqnarray}
\overline{B} = {{\int_{\lambda1}^{\lambda2} S(\lambda) \lambda B(\lambda) F(\lambda) d\lambda}
        \over{\int_{\lambda1}^{\lambda2} S(\lambda) \lambda F(\lambda) d\lambda}}
, \end{eqnarray}

\noindent
where S($\lambda$) is the response function  for each passband. In order to generate the average 
beaming factors, we adopted in Eq.~(3) the transmission curves for the Sloan ($u'g'r'i'z'y'$), 
UBVRI, HiPERCAM, Kepler, TESS, and Gaia passbands.  We did not consider reddening in the present calculations. 
We note   that Eq.~(3)  is suitable for photon-counting devices, such as CCDs. 

Figure 1 shows the comparison between beaming factors computed using Eq.~(3) (continuous line) and 
computed adopting the black body approach (stars) for HiPERCAM and using DA models at $\log g = 5.0$. The differences between the two 
recipes are notable, particularly for shorter effective wavelengths. Also noteworthy is the 
peak of beaming in the near-ultraviolet region around T$_{\rm eff}$ = 10\,000~K. This feature is also 
detected for less compact star models generated with the ATLAS9 code (see Fig.~6). 
Such behavior is related to the strength of the Balmer jump, whose effect is more intense for models with 
T$_{\rm eff}\approx$ 10\,000~K. On the other hand, if we assume that the flux is given by the Planck function,  
then for large values of  $\lambda$ and/or T$_{\rm eff}$ the calculations based on the black 
body approximation are closer to the predictions given by Eq.~(2) using realistic atmosphere models 
($\approx 1.0$), as expected.

\begin{figure}
        \includegraphics[height=9.cm,width=8.4cm,angle=0]{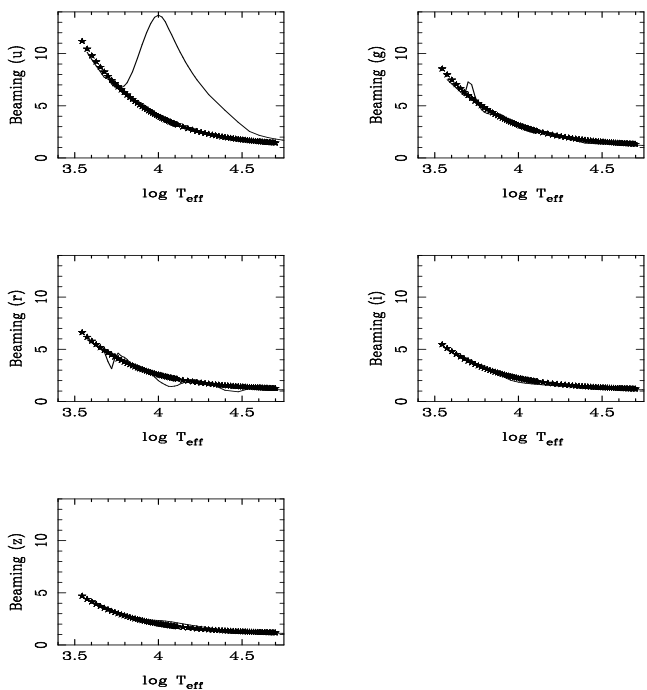}
        \caption{Comparison between the beaming factors  according to Eq.~(3) (HiPERCAM passbands). 
        The DA white dwarf models are represented by    
continuous lines  and  black body approximation by stars. All models are at $\log g$ = 5.0. 
}
\end{figure}

\begin{figure}
        \includegraphics[height=10.cm,width=8.4cm,angle=0]{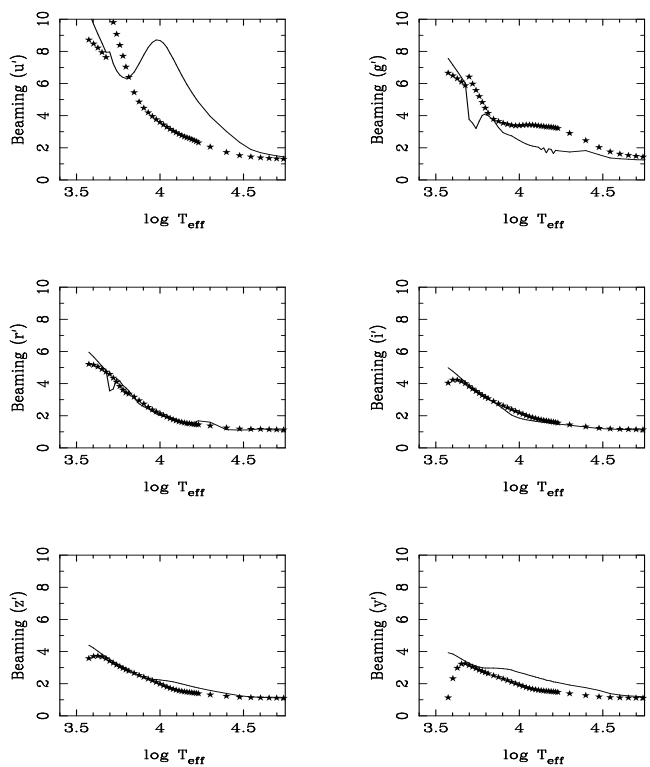}
        \caption{Comparison between the beaming factors (Sloan passbands, DA white dwarf models)  according to Eq.~(3). 
                Continuous lines indicate models with $\log g$ = 5.0 while  stars denote those with $\log g$ = 9.0.}
\end{figure}

Unlike the black body approach, the calculations that adopt more realistic models of stellar 
atmospheres  predict a dependence of the beaming factors on the local gravity, as can be seen 
in Fig. 2. The differences in beaming factors due to $\log g$ are more prominent for shorter wavelengths, although 
for the $y'$ passband there is also a notable difference for cooler models. 
For hot models and long effective wavelengths, the beaming factors are almost independent of $\log g$ 
because the models are close to the Rayleigh–Jeans law of a black body. 
The effects of local gravity on the beaming factors are very interesting, particularly for the 
Sloan ($u'$) filter. In general, the strength of the Balmer jump decreases with increasing local 
gravity of the white dwarf model (because of nonideal gas effects), and this impacts the beaming 
calculations for shorter effective wavelengths. The computation of the beaming factors depends on the 
strength of the Balmer jump, which 
in turn depends on the local gravity and effective temperature. We note that the Rayleigh-Jeans law 
and continuum opacities also contribute to the Balmer jump.

In Fig.~3, we show the influence of the hydrogen abundance in the beaming factor profiles.  The differences 
are small for longer effective wavelengths but  they can be detected 
observationally, particularly for effective temperatures below 20\,000 K and for shorter effective  wavelengths. 
As we anticipated at the beginning of this section, realistic atmosphere models for white dwarfs are able 
to predict a beaming factor dependence on the local gravity and with the hydrogen content, particularly     
for shorter effective wavelengths.

\begin{figure}
        \includegraphics[height=10.cm,width=8.4cm,angle=0]{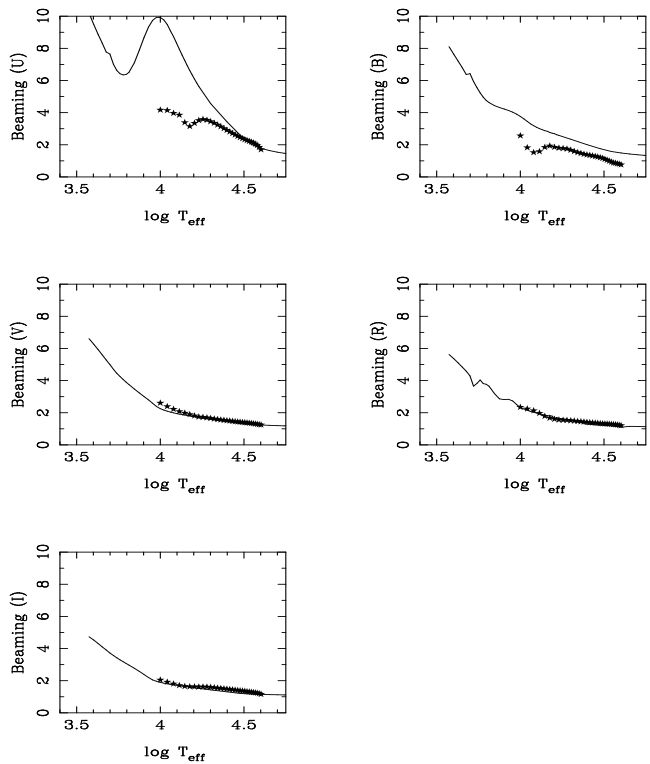}
        \caption{Beaming factors for the UBVRI passbands  according to Eq.~(3) for $\log g$ = 5.0.
                Continuous lines indicate DA white dwarf models while stars denote DB models.}
\end{figure}

\begin{figure}
        \includegraphics[height=10.cm,width=8.4cm,angle=0]{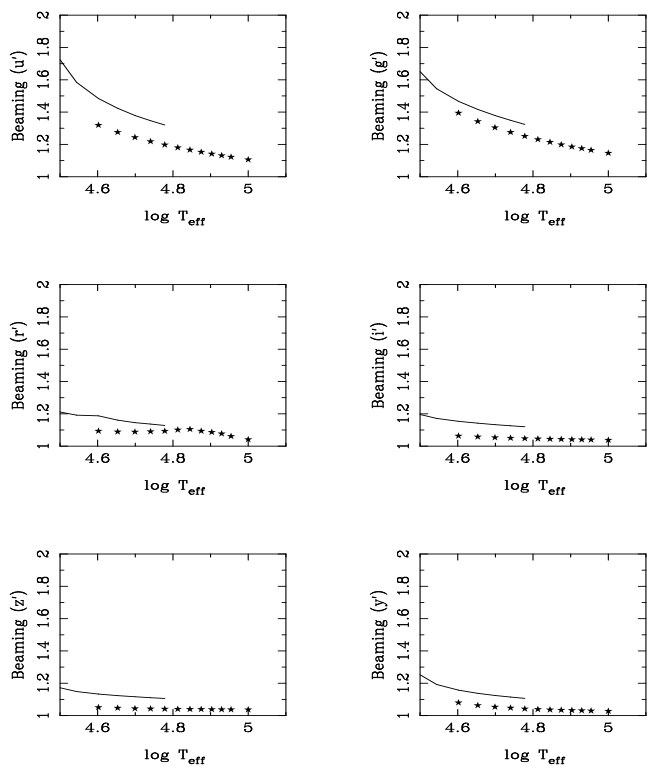}
        \caption{Beaming factors (Sloan passbands)  according to Eq.~(3) for log g=7.0.
                Continuous lines indicate LTE DA models and stars indicate NLTE DA models.}
\end{figure}

\begin{figure}
        \includegraphics[height=8.cm,width=8.cm,angle=0]{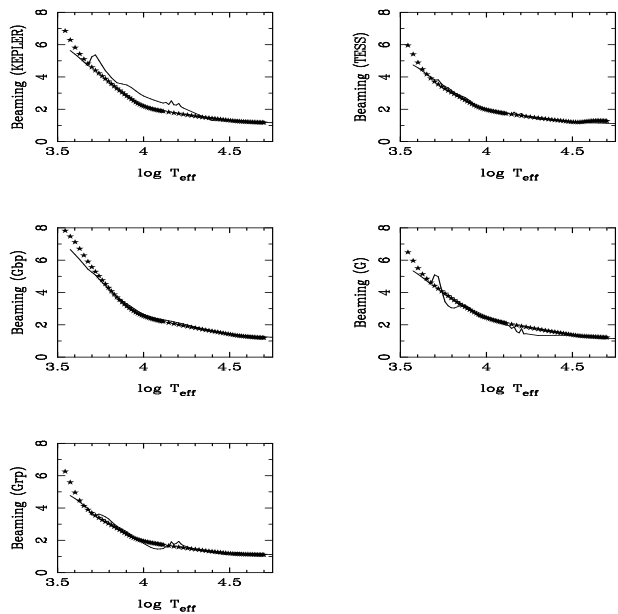}
        \caption{Beaming factors for main sequence ATLAS9 models (solar metallicity, 
stars) and DA models 
(continuous lines) according to Eq.~(3). All models are at $\log g$ =5.0 and shown 
for the Kepler, TESS and Gaia Gbp, G, and Grp passbands.}
\end{figure}

\begin{figure}
        \includegraphics[height=8.cm,width=9.0cm,angle=0]{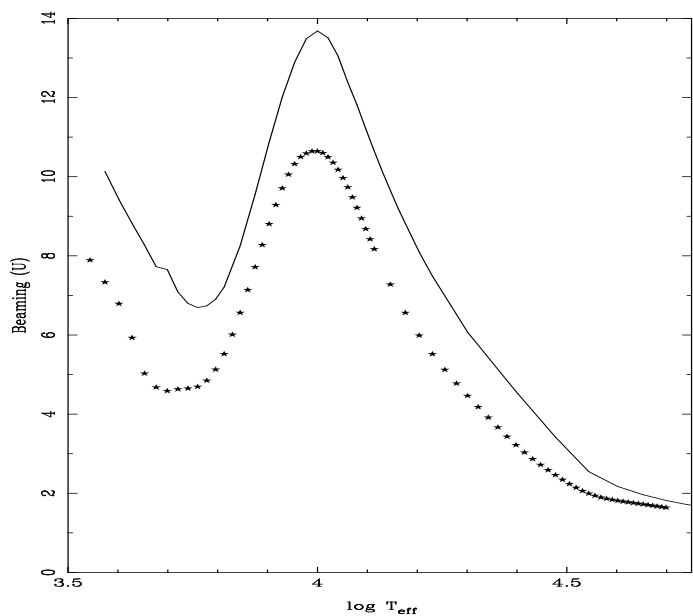}
        \caption{The same as in Fig. 5 but for the $u$ passband.}
\end{figure}

\begin{figure}
        \includegraphics[height=8.cm,width=9.0cm,angle=0]{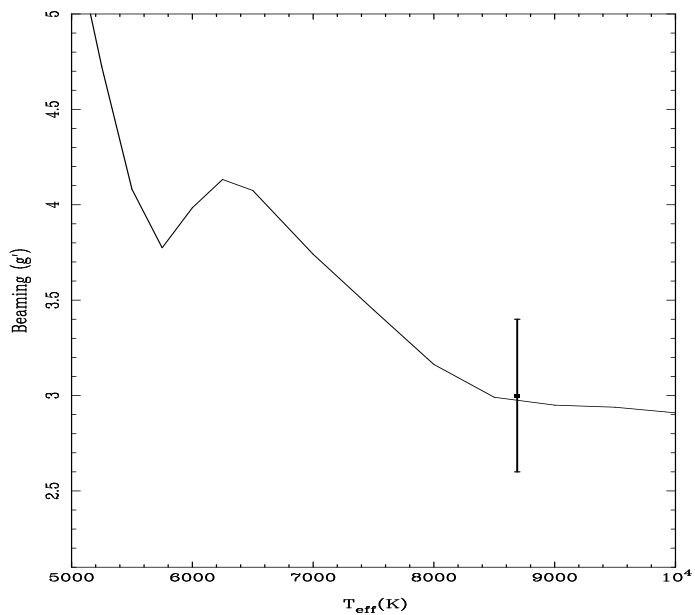}
        \caption{Comparison between the semi-empirical beaming factor (error bars) for the system NLTT\,11748 for the $g'$ passband and
        theoretical predictions adopting DA models at $\log g$ = 6.5.}
\end{figure}

Figure~4 illustrates the differences between the beaming factors for DA and DA-NLTE models for the Sloan passbands. 
The beaming factors computed for DA models are systematically larger than the corresponding DA-NLTE models for the 
entire instrumental range and for all considered effective temperatures. It is interesting to compare 
the systematic differences between the DA and DA-NLTE models with the results shown in Fig. 10 in Claret et al. (2020), 
where the effects of NLTE on the gravity-darkening coefficients (GDCs) are shown. From that figure, it was 
possible to determine the effective temperature limit for which it is necessary to adopt NLTE, 
around 40000 K, in agreement with the results by Gianninas et al. (2013), although this indicator  
cannot be considered as definitive. The "inefficiency" of the beaming factors in distinguishing this 
effective temperature limit    may be due to the fact that, in the case of GDC calculations,  we 
consider derivatives of central specific intensities with respect to Teff and log g (whose values are 
relatively small) while calculating the beaming factors; the derivatives $\left({d\ln F(\lambda)}\over{d\ln\lambda}\right)$ 
are  very much larger, especially at the Balmer jump and in the environment of spectral lines.  Another complementary 
possibility to explain the appearance of Fig. 4 is likely related to the use of different 
codes for local thermodynamic equilibrium (LTE) and NLTE. 

Concerning the ATLAS9 models, Fig.~5 shows the beaming factors for the Kepler, TESS, and Gaia passbands 
for models on the main sequence (solar metallicity, stars) and  DA white dwarf models,
denoted by continuous lines for a fixed value of $\log g$ = 5.0. 
In general, the differences are not very large  because we are dealing with relatively long  wavelengths. 
The situation changes for the ultraviolet range. In Fig.~6, we show the 
behavior of beaming factors of the same models as in Fig.~5 but for the $u$ filter;  the differences 
are notable in magnitude, in addition to being systematically offset. The systematic effects 
illustrated in Fig. 6 are due to the fact that the respective codes adopt different 
equations of state and helium content. 

 The comparison between theoretical and semi-empirical beaming factors is a difficult task mainly 
due to the  scarcity of observational data. Here we analyze the  semi-empirical beaming data 
related to white dwarfs  in the passband $g'$, published by Shporer et al.~(2010). 
These authors, analyzing the system NLTT\,11748, a non-interacting eclipsing double white 
dwarf binary with optical light dominated by an extremely low mass (ELM) DA white dwarf, 
determined $\overline{B(g')}$ = 3.0 $\pm$ 0.4. The system consists of He-core and C/O-core 
        white dwarfs but both atmospheres are pure hydrogen.  
We present in Fig.~7 an  exploratory comparison between 
the theoretical predictions for the passband $g'$ and the value inferred in Shporer et al. (2010). 
The  theoretical  beaming compares well with the derived semi-empirical value. 
On the other hand, Bloemen et al. (2011), studying the KPD 1946 + 4340 system, observationally 
determined $\overline {B (Kepler)}$  = 1.33 $\pm$0.02 for the sdB star. Since we did not 
calculate the beaming factors specifically for sdB-type stars, we used the results 
of DA and DB models to compare. Although this comparison is not entirely consistent, 
it can provide us with some clues. The respective values of the beaming factors are 
1.34 and 1.32 and are in agreement with the semi-empirical value given by Bloemen et al. (2011).

The beaming factor calculations for stars on the main sequence or the giant branch are also 
scarce in the literature. Bloemen et al.~(2012), for example, present the calculation of the beaming factor for 
an A-type main-sequence star  (KOI-74) in the  Kepler passband and find $\overline {B}$ = 2.22$\pm$0.04. 
An inspection in our Table~9 (see Table A.1) shows that the value of the beaming factor for the ATLAS9 model (T$_{\rm eff}$
 = 9500 K, $\log g$ = 4.3, and solar metallicity) is around 2.21, in  good agreement with both 
 the value mentioned above and the semi-empirical value (2.24$\pm$0.05).

\section{Limb-darkening coefficients for  DA-3D and DB-3D models}

Claret et al.~2020 derived LDCs for a set of one-dimensional DA, DB, and DBA white 
dwarf models. Here we compare these results to 3D model atmospheres for DA white dwarfs (DA-3D; 
Tremblay et al. 2013) in the ranges $\log g$ = 7.0-9.0 and T$_{\rm eff}$ = 6000-15\,000\,K, and for DB/DBA 
white dwarfs (DBA-3D; Cukanovaite et al. 2018, 2019) in the ranges $\log g$ = 7.5-9.0 and 
T$_{\rm eff}$ = 12\,000-34\,000\,K.  As usual, we adopted the following   limb-darkening laws  for 
the DA-3D and DB-3D models: 
the linear law (Schwarzschild 1906, Russell 1912, Milne 1921)
\begin{eqnarray}
          \frac{I(\mu)}{ I(1)} = 1 - u(1 - \mu),  
\end{eqnarray}

\noindent
the quadratic law (Kopal 1950)

\begin{eqnarray}
        \frac{I(\mu)}{ I(1)} = 1 - a(1 - \mu) - b(1 - \mu)^2,
\end{eqnarray}

\noindent
the square-root law (Díaz-Cordovés \& Giménez 1992)

\begin{eqnarray}
      \frac{I(\mu)}{ I(1)} =  1 - c(1 - \mu) - d(1 - \sqrt{\mu}),
\end{eqnarray}

\noindent
the logarithmic law (Klinglesmith \& Sobieski 1970)

\begin{eqnarray}
\frac{I(\mu)}{ I(1)} =  1 - e(1 - \mu) - f\mu\ln(\mu),
\end{eqnarray}

\noindent
the power-2 law  (Hestroffer 1997) 

\begin{eqnarray}
\frac{I(\mu)}{ I(1)} =  1 - g(1 - \mu^h),
\end{eqnarray}

\noindent
and a four-term law (Claret 2000) 

\begin{eqnarray}
\frac{I(\mu)}{ I(1)} = 1 - \sum_{k=1}^{4} {a_k} (1 - \mu^{\frac{k}{2}}), 
\end{eqnarray}

\noindent
where  $I(1)$ is the specific intensity at the center of the disk and 
 $u, a, b, c, d, e, f, g, h$, and $a_k$ are the corresponding LDCs. 
 For the calculation of the LDCs, we adopted the least-squares method (LSM). For details regarding 
 this choice, see Claret et al. (2020).

The merit function for each law and passband is  given by

\begin{eqnarray}
{\chi^2}= \sum_{i=1}^{N} \left( {y_i - Y_i}\right)^2
,\end{eqnarray}

\noindent
where $y_i$ is the model intensity at point $i$, $Y_i$ is the fitted function at the same 
point, and $N$ is the number of $\mu$ points. The $\chi^2$ are given in Tables~11-34 
 (see Appendix A) for user orientation for each law and passband.

The advantages of the four-term law over all bi-parametric or linear laws are widely discussed 
in Claret (2000) for main sequence and giant stars, and for the case of white dwarfs in  
Claret et al.~(2020). In the case of 3D models for main sequence stars and giants, a  
discussion on the superiority of the four-term law can be found in Magic et al.~(2015). 
In the case of the DA-3D and DB-3D models, the corresponding difference in $\chi^2$ between
the six laws gives similar results to those obtained in the aforementioned papers.

\begin{figure}
\includegraphics[height=8.cm,width=9.0cm,angle=0]{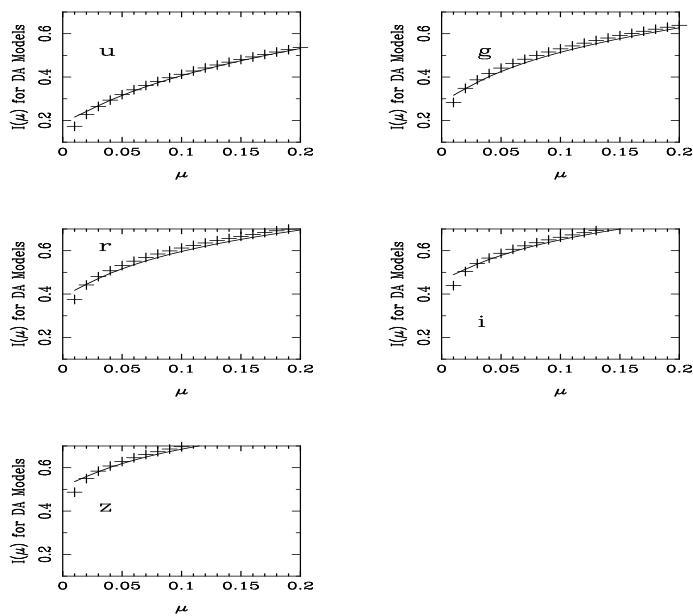}
\caption{Distribution of specific intensities for 1D (continuous line) and 3D 
(crosses) DA models with T${_{\rm eff}}$ = 6000 K and $\log g$ = 8.0 for the HiPERCAM instrumental system.}
\end{figure}

\begin{figure}
\includegraphics[height=8.cm,width=9.0cm,angle=0]{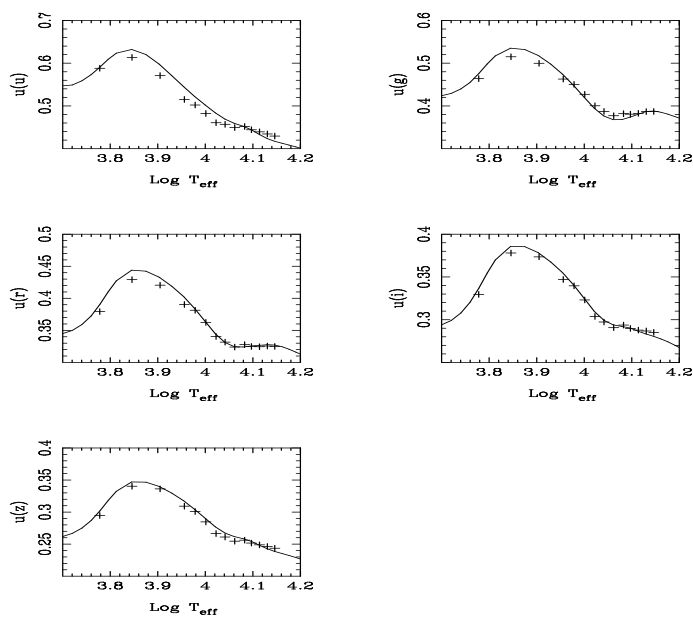}
\caption{Comparison between linear LDCs for HiPERCAM  passbands for DA-3D models (crosses) and 
        DA-1D models (continuous line) at $\log g$ = 7.0.}
\end{figure}

\begin{figure}
        \includegraphics[height=8.cm,width=9.0cm,angle=0]{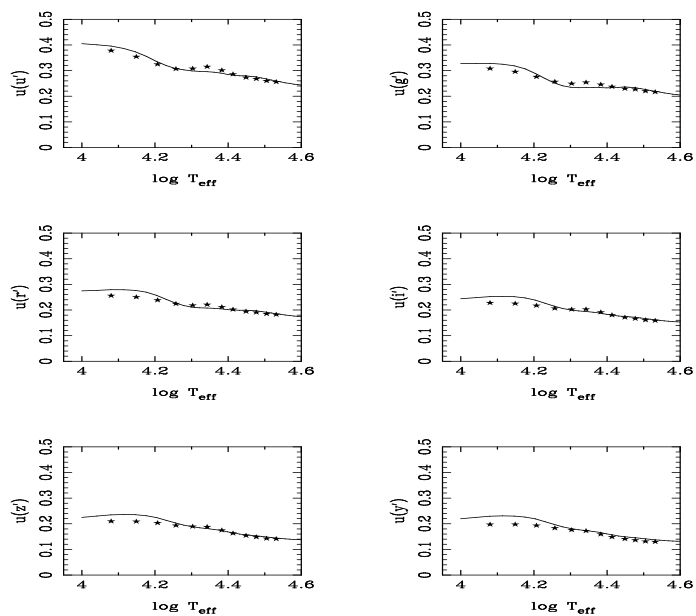}
        \caption{Comparison between linear LDCs and Sloan  passbands for DB-3D models (stars) and 
                DB-1D models (continuous line) at $\log g  =$ 8.0. }
\end{figure}

In our previous paper on LDCs and GDCs for white dwarfs, we only investigated 1D models. 
Our main objective in this section is to study the effects of 3D models on LDCs and GDCs. 
However, for the calculation of the GDCs, we need to calculate the term $\left(\frac{\partial{\ln I_o(\lambda)}}
{\partial{\ln g}}\right)_{T_{\rm eff}}$, where  $I_o(\lambda)$ is the specific 
intensity at a given 
wavelength  at the center of the stellar disk and the subscript  T${_{\rm eff}}$ denotes 
a  derivative  at constant effective temperature. The impact of this 
contribution to the GDC is not very large, but it is not negligible. Unfortunately, 
the present 3D  grids are not  regular with respect to effective temperatures, 
which prevents us from performing consistent calculations of GDCs.  We will address these 
calculations in the future, since they require an enormous amount of personal and computational time. 
However, we can calculate the GDCs for an individual system  upon request from interested parties.  

Before analyzing the LDCs for the white dwarfs, we will note that a very important  but sometimes overlooked point 
 is that a close binary system should not be treated as if each star in the pair behaves as an 
isolated star, since the proximity effects can change the individual properties of each star. For example, mutual irradiation between the components alters the spectra, occasionally
resulting in a very different spectrum from a non-irradiated white dwarf. This can alter 
the distribution of specific intensities (limb-darkening), effective temperatures, 
and even the observed chemical compositions. In many cases, this type of interaction can lead to 
erroneous determinations of the stellar parameters  as indicated by  Claret \& Gim\'enez (1990, 1992) 
and Claret (2001, 2004, 2007). In this section, we work with the hypothesis that 
the stars in a close binary system behave like isolated stars, but we plan to implement 
the proximity effects in the future.

In Fig.~8, we compare the behavior of the specific intensities as a function 
of $\mu$ for the one- and three-dimensional DA models with T$_{\rm eff}$ = 6000 K and log g = 8.0 (HiPERCAM). 
We note that the effective temperature is not exactly the same as the T$_{\rm eff}$ of the 3D model, which 
is 5998 K. However, this difference is essentially negligible for comparison purposes. 
Examining Fig.~8, we can see a general trend that the 1D model is brighter than the corresponding 
3D model toward the limb for all passbands of the HiPERCAM instrumental system. 
Such an effect has been previously noted by Magic et al.~(2015) when these authors compared 
the distribution of intensities for the ATLAS (1D) and Stagger (3D) models of main sequence stars. 
We also find that this effect depends on the local gravity: the larger the $\log g$, 
the smaller the effect, provided the other parameters of the input physics are fixed. 
An explanation for this effect may be that the temperature stratification 
and temperature gradients are different in the 1D and 3D modeling (Magic et al.~2015). 
One contribution that changes the temperature stratification for DA and DB white dwarfs 
is the convective overshoot seen in 3D models, which cools the upper layers 
(Tremblay et al.~2013; Cukanovaite et al.~2018). Since regions toward the limb have a 
longer optical path across the overshoot region, they become less bright in 3D.

On the other hand, as is widely known, the linear approximation for limb-darkening does not correctly 
describe the distribution of specific intensities. However, this law allows us to 
more easily compare the results from  models with different input physics. 
 We will use it here for the purpose of studying the 
direct effects of 3D on LDCs. Comparisons of the  LDCs between DA-1D and DA-3D models 
are shown in Fig.~9  (HiPERCAM). Significant differences can only be distinguished for shorter 
effective wavelengths and lower effective temperatures,  and these could be at the limit of observational  
detection. Three-dimensional effects due to convective energy transport become significant at 
low T$_{\rm eff}$ and thus the differences between 1D and 3D at these temperatures are not unexpected.

In Fig.~10, we compare the linear LDCs for DB-3D models (stars) with their 1D counterparts. 
There is a systematic effect in the sense that the LDCs of the 3D models are smaller for 
T$_{\rm eff}  <$ 20\,000 K. Similarly to DA white dwarfs, 3D effects become significant 
in this temperature regime (Cukanovaite et al.~2018).

According to the results shown in Fig.~8, the distribution of specific intensities 
for 3D models indicates that, in general, these models are less bright than their 1D counterparts, 
which implies steeper profiles toward the limb. To better describe these intensities,
we recommend the use of the four-term law given the level of precision that is being  achieved 
with Earth-based instruments as well as space missions, such as Kepler, TESS, and, in the near future, PLATO.

\section{Summary}

The beaming factor calculations presented here, based on more 
realistic atmosphere models,  clearly indicate 
that the black body approximation is not accurate. This is particularly 
important   for the filters $u$, $u'$, $U$, $g$, $g'$, and $B$ since the 
black body approach is only valid for high effective temperatures 
and/or long effective wavelengths. 
 To check the validity of our calculations of beaming factors, we compared 
 them with the few existing cases in the literature (theoretical and semi-empirical), 
 finding a very good agreement. 
 To more accurately analyze the 
light curves of white dwarfs, we recommend the use of the beaming factors presented here. 
On the other hand, concerning limb-darkening, we recommend 
the use of the  3D  four-term law since    
 Hayek et al. (2012) has shown that 3D models lead to a better fit for the HD 189733 system
when compared to the predictions of the 1D models. 
 Finally, Tables A1 and A2 in the appendix give information 
on the data available at the CDS (Centre de Données Astronomiques de Strasbourg) 
or directly from the authors.

\begin{acknowledgements} 
         We thank the referee S. Bloemen for his helpful suggestions and comments. 
 The Spanish MEC (ESP2017-87676-C5-2-R,  PID2019-107061GB-C64, and  
 PID2019-109522GB-C52) is gratefully acknowledged for its 
support during the development of this work. A.C. also 
acknowledges financial support from the State Agency for 
Research of the Spanish MCIU through the “Center of 
Excellence Severo Ochoa” award for the Instituto de 
Astrofísica de Andalucía (SEV-2017-0709). The research leading to 
these results has received funding from the European Research 
Council under the European Union's Horizon 2020 research and 
innovation programme n. 677706 (WD3D). SGP acknowledges the 
support of a Science and Technology Facilities Council (STFC) 
Ernest Rutherford Fellowship. This research has made 
use of the SIMBAD database, operated at the CDS, Strasbourg, 
France, and of NASA's Astrophysics Data System Abstract Service.
\end{acknowledgements}

{}

\begin{appendix}

\section{Brief description of Tables A1 and A2}

Tables A1 and  A2  summarize the types 
of data available as well as the central intensities for each 
photometric system  in ergs/cm$^2$/s/Hz/ster in the case of LDCs for  the 3D models. 
 For more details, see the ReadMe file 
at the CDS website.

\renewcommand{\tablename}{Table }       
\begin{table*}  

\caption{Doppler beaming {\bf factors} for the Kepler, Tess, Gaia, HiPERCAM, Sloan, 
        and UBVRI photometric systems.}
\begin{flushleft}
\begin{tabular}{lcccccclc}   
                                      
\hline                         
 Name    & Source   &  range T$_{\rm eff}$ & range $\log g$ & $\log$ [H/He] &Beaming/Filters   \\ 
                        \hline   
Table1 &{\sc DA}      &3750 K-60000 K  & 5.0-9.5&  0.00                      &{Kepler/TESS/Gaia/HiPERCAM}  \\
Table2 &{\sc DB}      &10000 K-40000 K & 5.5-9.5&  $-$10.0                     &{Kepler/TESS/Gaia/HiPERCAM}  \\
Table3 &{\sc DBA}     &11000 K-40000 K & 7.0-9.0&  $-$2.0                      &{Kepler/TESS/Gaia/HiPERCAM}  \\
Table4 &{\sc DA-NLTE} &40000 K-100000 K& 6.0-9.5&  0.0                       &{Kepler/TESS/Gaia/HiPERCAM}  \\
Table5 &{\sc DA}      &3750 K-60000 K  & 5.0-9.5&  0.0                       &{sloan/UBVRI}  \\
Table6 &{\sc DB}      &10000 K-40000 K & 5.5-9.5&  $-$10.0                     &{sloan/UBVRI}  \\
Table7 &{\sc DBA}     &11000 K-40000 K & 7.0-9.0&  $-$2.0                      &{sloan/UBVRI}  \\
Table8 &{\sc DA-NLTE} &40000 K-100000 K& 6.0-9.5&  0.0                       &{sloan/UBVRI}  \\
Table9 &{\sc ATLAS9}  &3500 K-50000 K  & 0.0-5.0&  $-$2.5-0.5\tablefootmark{a} &{Kepler/TESS/Gaia/HiPERCAM}   \\
Table10 &{\sc ATLAS9} &3500 K-50000 K  & 0.0-5.0&  $-$2.5-0.5\tablefootmark{a} &{sloan/UBVRI}   \\
\hline
\hline
\end{tabular}
\tablefoot{
\tablefoottext{a}{Values of  metallicities for the ATLAS9 models}}
\end{flushleft}
\end{table*}

\renewcommand{\tablename}{Table }       
\begin{table*}  
        
\caption{Limb-darkening coefficients for the Kepler, Tess, Gaia, HiPERCAM, Sloan, and UBVRI photometric systems.}
\begin{flushleft}
\begin{tabular}{lcccccclc}                         
\hline                         
Name    & Source   &  range T$_{\rm eff}$ & range log $g$ & log [H/He] & Filters & Fit/equation   \\ 
\hline   
Table11 &{\sc DA-3D}  &5998 K- 14986K & 7.0-9.0&  0.0 &{Kepler/TESS/Gaia/HiPERCAM} & LSM/Eq. 4 \\
Table12 &{\sc DA-3D}  &5998 K- 14986K & 7.0-9.0&  0.0 &{Kepler/TESS/Gaia/HiPERCAM} & LSM/Eq. 5 \\
Table13 &{\sc DA-3D}  &5998 K- 14986K & 7.0-9.0&  0.0 &{Kepler/TESS/Gaia/HiPERCAM} & LSM/Eq. 6 \\
Table14 &{\sc DA-3D}  &5998 K- 14986K & 7.0-9.0&  0.0 &{Kepler/TESS/Gaia/HiPERCAM} & LSM/Eq. 7 \\
Table15 &{\sc DA-3D}  &5998 K- 14986K & 7.0-9.0&  0.0 &{Kepler/TESS/Gaia/HiPERCAM} & LSM/Eq. 8 \\
Table16 &{\sc DA-3D}  &5998 K- 14986K & 7.0-9.0&  0.0 &{Kepler/TESS/Gaia/HiPERCAM} & LSM/Eq. 9 \\
Table17 &{\sc DA-3D}  &5998 K- 14986K & 7.0-9.0&  0.0 &{Sloan/UBVRI} & LSM/Eq. 4 \\
Table18 &{\sc DA-3D}  &5998 K- 14986K & 7.0-9.0&  0.0 &{Sloan/UBVRI} & LSM/Eq. 5 \\
Table19 &{\sc DA-3D}  &5998 K- 14986K & 7.0-9.0&  0.0 &{Sloan/UBVRI} & LSM/Eq. 6 \\
Table20 &{\sc DA-3D}  &5998 K- 14986K & 7.0-9.0&  0.0 &{Sloan/UBVRI} & LSM/Eq. 7 \\
Table21 &{\sc DA-3D}  &5998 K- 14986K & 7.0-9.0&  0.0 &{Sloan/UBVRI} & LSM/Eq. 8 \\
Table22 &{\sc DA-3D}  &5998 K- 14986K & 7.0-9.0&  0.0 &{Sloan/UBVRI} & LSM/Eq. 9 \\
Table23 &{\sc DB-3D}  &12098 K- 34105 K& 7.5-9.0&  $-$10.0 &{Kepler/TESS/Gaia/HiPERCAM} & LSM/Eq. 4 \\
Table24 &{\sc DB-3D}  &12098 K- 34105 K& 7.5-9.0&  $-$10.0 &{Kepler/TESS/Gaia/HiPERCAM} & LSM/Eq. 5 \\
Table25 &{\sc DB-3D}  &12098 K- 34105 K& 7.5-9.0&  $-$10.0 &{Kepler/TESS/Gaia/HiPERCAM} & LSM/Eq. 6 \\
Table26 &{\sc DB-3D}  &12098 K- 34105 K& 7.5-9.0&  $-$10.0 &{Kepler/TESS/Gaia/HiPERCAM} & LSM/Eq. 7 \\
Table27 &{\sc DB-3D}  &12098 K- 34105 K& 7.5-9.0&  $-$10.0 &{Kepler/TESS/Gaia/HiPERCAM} & LSM/Eq. 8 \\
Table28 &{\sc DB-3D}  &12098 K- 34105 K& 7.5-9.0&  $-$10.0 &{Kepler/TESS/Gaia/HiPERCAM} & LSM/Eq. 9 \\
Table29 &{\sc DB-3D}  &12098 K- 34105 K& 7.5-9.0&  $-$10.0 &{Sloan/UBVRI} & LSM/Eq. 4 \\
Table30 &{\sc DB-3D}  &12098 K- 34105 K& 7.5-9.0&  $-$10.0 &{Sloan/UBVRI} & LSM/Eq. 5 \\
Table31 &{\sc DB-3D}  &12098 K- 34105 K& 7.5-9.0&  $-$10.0 &{Sloan/UBVRI} & LSM/Eq. 6 \\
Table32 &{\sc DB-3D}  &12098 K- 34105 K& 7.5-9.0&  $-$10.0 &{Sloan/UBVRI} & LSM/Eq. 7 \\
Table33 &{\sc DB-3D}  &12098 K- 34105 K& 7.5-9.0&  $-$10.0 &{Sloan/UBVRI} & LSM/Eq. 8 \\
Table34 &{\sc DB-3D}  &12098 K- 34105 K& 7.5-9.0&  $-$10.0 &{Sloan/UBVRI} & LSM/Eq. 9 \\
\hline
\hline
\end{tabular}
\end{flushleft}
\end{table*}

\end{appendix}

\end{document}